# Quantum gate algorithm for reference-guided DNA sequence alignment

G. D. Varsamis[1], I. G. Karafyllidis[1,2]*, K. M. Gilkes[3], U. Arranz[3], R. Martin-Cuevas[3], G. Calleja[3], P. Dimitrakis[2], P. Kolovos[4], R. Sandaltzopoulos[4], H. C. Jessen[3], J. Wong[3].

1 Department of Electrical and Computer Engineering, Democritus University of Thrace, Xanthi, 67100 Greece
2 National Centre for Scientific Research Demokritos, Athens, 15342 Greece
3. EY Global Innovation Quantum Computing Lab
4. Department of Molecular Biology and Genetics, Democritus University of Thrace, Alexandroupolis, 68100 Greece
*Corresponding author: *ykar@ee.duth.gr*

ABSTRACT
Reference-guided DNA sequencing and alignment is an important process in computational molecular biology. The amount of DNA data grows very fast, and many new genomes are waiting to be sequenced while millions of private genomes need to be re-sequenced. Each human genome has 3.2B base pairs, and each one could be stored with 2 bits of information, so one human genome would take 6.4B bits or ~760MB of storage ("National Institute of General Medical Sciences," n.d.). Today's most powerful tensor processing units cannot handle the volume of DNA data necessitating a major leap in computing power. It is, therefore, important to investigate the usefulness of quantum computers in genomic data analysis, especially in DNA sequence alignment. Quantum computers are expected to be involved in DNA sequencing, initially as parts of classical systems, acting as quantum accelerators. The number of available qubits is increasing annually, and future quantum computers could conduct DNA sequencing, taking the place of classical computing systems. We present a novel quantum algorithm for reference-guided DNA sequence alignment modeled with gate-based quantum computing. The algorithm is scalable, can be integrated into existing classical DNA sequencing systems and is intentionally structured to limit computational errors. The quantum algorithm has been tested using the quantum processing units and simulators provided by IBM Quantum, and its correctness has been confirmed.

## 1. Introduction

Genomics is a major generator of Big Data and is highly demanding in data acquisition, storage, distribution and analysis (Stephens et al., 2015). Genomic data is growing, and it is expected to grow exponentially in the near future (Camarillo-Guerrero et al., 2021; Edgar et al., 2022). Reference-guided DNA sequencing and alignment is an essential step upon next generation sequencing and in many personalized medicine applications, such as CRISPR-Cas9 screening, which generates large amounts of genomic data in an attempt to identify novel target genes for therapeutic treatment (Krittanawong et al., 2017; Ponting, 2017). Quantum computers use quantum mechanical properties and can handle Big Data more efficiently than their classical counterparts, as the number of the available qubits increases (Shaikh and Ali, 2016). It is, therefore, important to investigate the usefulness of quantum computers in genomic data analysis, especially in DNA sequence alignment.

Quantum gate algorithms for DNA sequence alignment have been developed and showed a promising prospective in handling the amount of data generated in this process (De, 2022; Emani et al., 2021; Hollenberg, 2000; Pattel, A. Rathee et a., 2021). These algorithms are based on Grover's quantum search algorithm and its variations (Grover, 1996). This approach poses three significant problems. First, the quantum gate circuits that implement these algorithms are complicated, have large circuit depths and demand a very large number of qubits, which makes them prominent to computational errors. Secondly, although the quantum algorithms based on Grover search may be effective in the future when larger quantum computers will be available, they cannot be incorporated into today's classical DNA sequencing systems to form an effective hybrid classical/quantum computing system to obtain a significant acceleration of DNA sequence alignment. The third problem is that although these algorithms are effective with toy-model DNA sequences, they cannot be scaled easily, as the number of available qubits increases, to address real problems.

The purpose of reference-guided DNA sequence alignment is to compare and measure the similarity between one or several DNA segments–called sequencing Reads–and a (larger) reference DNA sequence or genome. The process aims to locate the region(s) in the reference DNA that is (are) identical or most similar. Many conclusions can be drawn from the results of the comparison, such as the occurrence of mutations and evolutionary relations (Berger et al., 2020; Li and Homer, 2010;).

We developed, and present here, a novel quantum algorithm for reference-guided DNA sequence alignment. Our algorithm has been developed using the gate or circuit model of quantum computation (Nielsen and Chuang, 2011; Karafyllidis, 2005). Quantum gate algorithms capture the details of biological processes and allow to form a connection between quantum effects in biology and quantum mechanics (Mohseni et al., 2014; Karafyllidis, 2008; Karafyllidis, 2012). Our quantum algorithm has three advantageous characteristics. First, it can be easily integrated into existing classical systems for DNA sequence alignment to act as a quantum accelerator. Secondly, the algorithm's quantum circuit is not prone to errors because it is shallow, and we avoided as much as possible the use of quantum phase gates. Thirdly, the scalability of the algorithm is straightforward, and it can handle easily larger DNA Reads and reference DNA sequences, as the number of available qubits increases. We tested and validated our quantum algorithm by simulating its operation using the software development kit Qiskit, and various quantum processing units and simulators provided by IBM Quantum. Finally, we describe a method to set the Read and reference DNA states in superposition, allowing the quantum computer to act as a quantum accelerator.

## 2. Outline of classical DNA alignment based on Hamming distance

Classical methods of DNA sequence alignment search for similarities between two sequences. The basic method for global alignment is the Needleman–Wunsch algorithm (Needleman and Wunsch, 1970) and the basic method for local alignment is the Smith–Waterman algorithm (Smith and Waterman, 1981). Since then, several alignment methods have been developed for both local and global alignment (Wilton and Szalay, 2022; Chao et al., 2022). The operations used in most of these algorithms is the compare and shift operations.

We formulated the two operations as follows. DNA sequences are usually represented as character strings, each character being a member of the DNA residues type set {A, C, G, T}. The reference DNA sequences R and the Read sequence Q that are to be compared are represented by two sets:

$$R = \{r_1, r_2, \cdots, r_m\} \qquad (1.1)$$

and

$$Q = \{q_1, q, \cdots, q_n\} \qquad (1.2)$$

In the above, equations $r$ represents a reference residue and $q$ a Read residue. The reference is much larger than the Read: $m \gg n$. The Read is compared with a part of the reference starting with a residue $r_k$ and ending with the residue $r_{k+n-1}$, where $n$ is the length of the Read. The result of the comparison is the Hamming distance, $H_k$, between the two strings:

$$H_k = \sum_{i=1}^{n} (q_i, r_{k+i-1}) \qquad (2)$$

The parentheses in equation (2) take the value "0" if the two residues are the same and the value "1" if the two residues differ. The index "k" in $H_k$ indicates the Hamming distance between the Read and the part of the reference that starts with $r_k$. If all residues match $H_k=0$ and if all residues differ $H_k=n$. Therefore, for partial residue matches, $0 < H_k < n$. The Hamming distance $H_k$ is stored, the Read is shifted by one residue, i.e. $r_{k+1}$, and the new Hamming distance, $H_{k+1}$, is computed:

$$H_{k+1} = \sum_{i=1}^{n}(q_i, r_{(k+1)+i-1}) \tag{3}$$

At the end of the computation a set of Hamming distances is created, and the Read is aligned at the reference site, where the Hamming distance is minimum:

$$Align = min\{H_k, H_{k+1}, H_{k+2}, \ldots\} \tag{4}$$

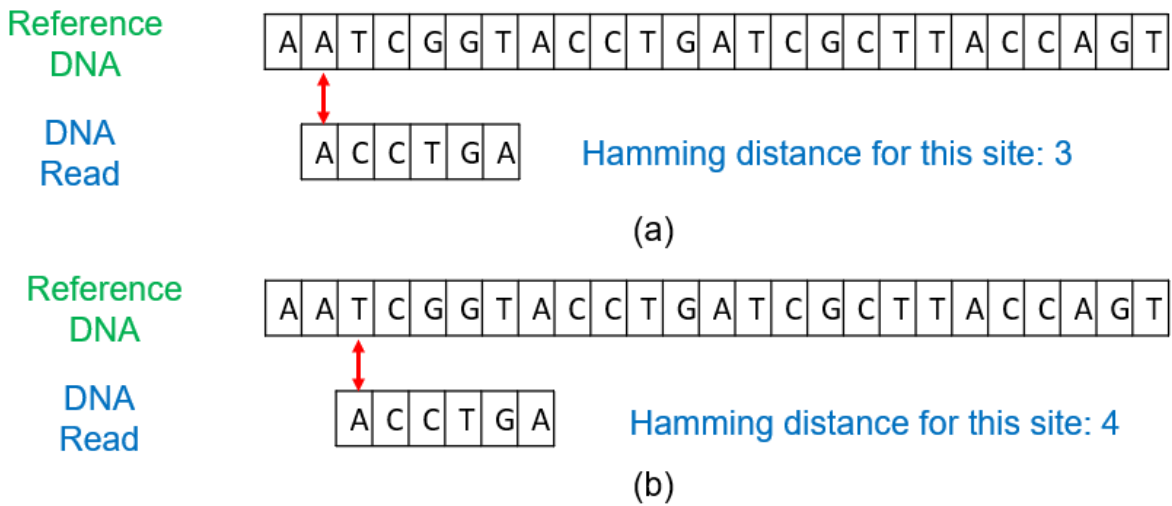

Figure 1. Schematic depiction of the classical compare and shift operations. (a) The Hamming distance is 3. (b) The Read is shifted by one reference site and the new Hamming distance becomes 4.

Figure 1 shows schematically the classical compare and shift operations. In Figure 1(a) the Read is located at the part of the reference that begins with the second residue. The Hamming distance in this case is 3. In Figure 1(b) the Read is shifted by one reference site and the Hamming distance becomes equal to 4.

## 3. Quantization of the classical algorithm: Compare and shift operators

Our aim is to develop a quantum algorithm that can be integrated into existing systems for DNA sequence alignment that use classical algorithms and computers, so that it acts as a quantum accelerator. To achieve this, we quantize the compare and shift operators. The quantized operators are expressed as quantum circuits that can be readily executed on a quantum computer. We use two qubits to encode each of the four residues. The qubit state labeling is as follows:

$$|A\rangle \to |00\rangle, \quad |T\rangle \to |11\rangle, \quad |C\rangle \to |01\rangle, \quad |G\rangle \to |10\rangle \quad (5)$$

*3.1 Quantum compare operator.*

Figure 2 shows the quantum circuit of the compare operation. The quantum circuit represents the "compare" operator, i.e., a unitary operator acting on the qubit states. The two qubits encoding a reference residue are labeled as $D_0$ and $D_1$. For example, $|C\rangle \to |D_0 D_1\rangle \to |01\rangle$. The two qubits encoding a Read residue are labeled as $R_0$ and $R_1$. For example, $|G\rangle \to |R_0 R_1\rangle \to |10\rangle$. The main obstacle in comparing qubit states comes from the no-cloning theorem, which states that a qubit state cannot be copied. To compare the qubits without copying we use CNOT quantum gates and an ancilla qubit, which is set to state $|0\rangle$.

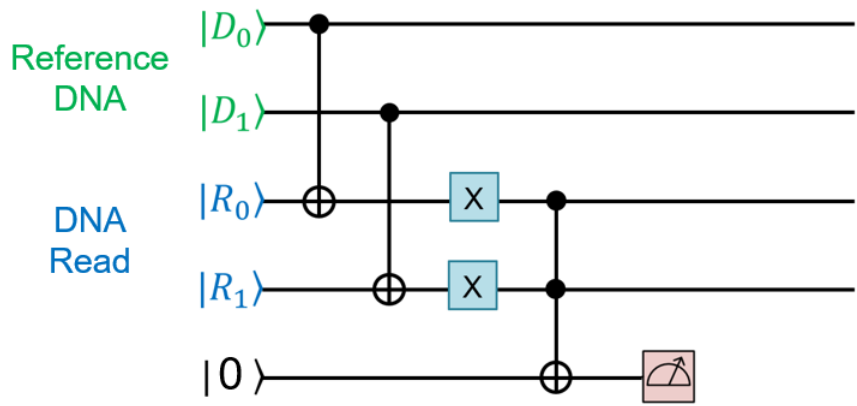

*Figure 2. The quantum circuit for the comparison operator. The ancilla qubit is set to state $|0\rangle$.*

The reference and Read qubits along with the ancilla qubit form a five-qubit quantum register, the initial state, $|q_1\rangle$, of which is:

$$|q_1\rangle = |D_0 \, D_1 \, R_0 \, R_1 \, 0 \rangle \tag{6}$$

Then, the two CNOT quantum gates act. If the reference and Read residues are the same the quantum register state, $|q_2\rangle$, becomes:

$$|q_2\rangle = |D_0 \, D_1 \, 0 \, 0 \, 0 \rangle \tag{7.1}$$

If the reference and Read residues differ, the quantum register will be in one of the following three states:

$$|q_2\rangle = |D_0 \, D_1 \, 0 \, 1 \, 0 \rangle \tag{7.2}$$
$$|q_2\rangle = |D_0 \, D_1 \, 1 \, 0 \, 0 \rangle \tag{7.3}$$
$$|q_2\rangle = |D_0 \, D_1 \, 1 \, 1 \, 0 \rangle \tag{7.4}$$

Next, two quantum X gates act on the Read qubits and flip their state. If the reference and Read residues are the same, the new quantum register state, $|q_3\rangle$, is:

$$|q_3\rangle = |D_0 \, D_1 \, 1 \, 1 \, 0 \rangle \tag{8.1}$$

In correspondence with equations (7.2), (7.3) and (7.4), if the reference and Read residues differ, the quantum register will be in one of the following three states:

$$|q_3\rangle = |D_0 \, D_1 \, 1 \, 0 \, 0 \rangle \tag{8.2}$$
$$|q_3\rangle = |D_0 \, D_1 \, 0 \, 1 \, 0 \rangle \tag{8.3}$$
$$|q_3\rangle = |D_0 \, D_1 \, 0 \, 0 \, 0 \rangle \tag{8.4}$$

Then a CCNOT quantum gate acts with the ancilla qubit as target. Both control qubits must be in state $|1\rangle$ to flip the state of the ancilla qubit from $|0\rangle$ to $|1\rangle$. This happens only for the case of equation (8.1) in which the Read and reference residues are the same. If they differ the ancilla qubit remains in state $|0\rangle$. At the end of the quantum computation of Figure (2), the ancilla qubit is measured. If the measurement results in state $|1\rangle$, the reference and Read residues are the same and if in state $|0\rangle$, the residues differ.

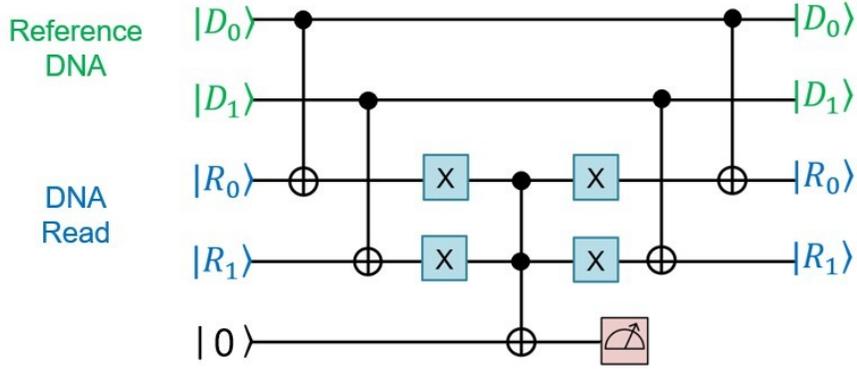

*Figure 3. The final quantum circuit for the comparison operator. The Read residue qubits are restored to their initial states.*

During the quantum computation of Figure 2, the states of the Read residue qubits changed. We use the time-reversibility of quantum computation, and we restore them to their initial states by applying the same quantum gates in reverse order to the four upper qubits.

The quantum circuit of Figure 3 compares the Read and reference residues by measuring the state of the ancilla qubit. The quantum circuit comprises CNOT and X quantum gates, which are less error prone than the phase shift and Hadamard quantum gates. The same circuit can be used to compare all the Read residues and send the values of the ancilla qubit to a classical computer, which will compute the Hamming distance for the specific site of the reference.

*3.2 Quantum shift operator.*
To compose the quantum circuit of the shift operation we use Swap quantum gates. A Swap gate is formed by applying three CNOT gates as shown in Figure 4. A Swap gate interchanges the states of the qubits $|x\rangle$ and $|y\rangle$ on which it acts.

Figure 5 shows a four-qubit quantum register. The two upper qubits $|R_0\rangle$ and $|R_1\rangle$ represent a Read residue. Two auxiliary qubits, both in state $|0\rangle$ are used. The initial state of the quantum register is $|R_0, R_1, 0, 0\rangle$. The two Swap gates act as shown in Figure 5 and interchange the qubit states. The final state of the quantum register is $|0, 0, R_0, R_1\rangle$, i.e., the Read state is shifted by one reference DNA site.

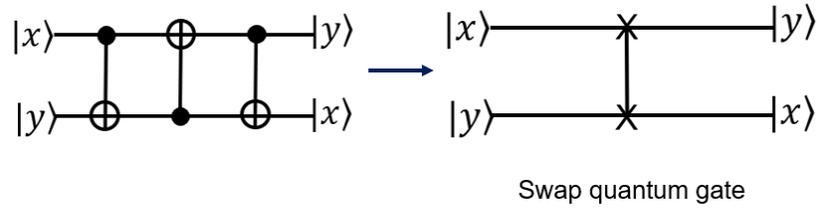

Figure 4. Three CNOT quantum gates form a Swap quantum gate.

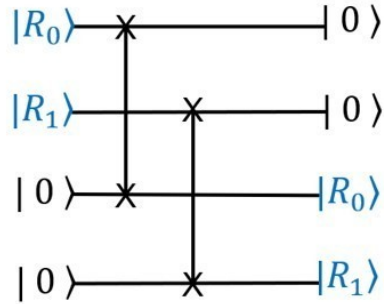

Figure 5. A Read residue is shifted using two Swap gates.

Next, we show how the entire DNA Read can be shifted by one reference DNA site using the basic circuit of Figure 5. We add one more index to the Read state labeling. The index indicates the site of residue in the Read sequence. For example, the two qubits encoding the Read residue at site $m$ are $|R_0^m\rangle$ and $|R_1^m\rangle$. Figure 6 shows the quantized shift operation, i.e., the shift operator that shifts the entire Read by one reference site.

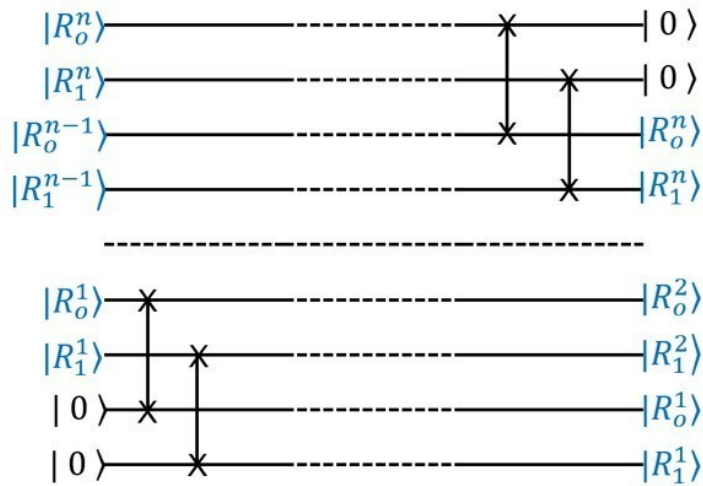

Figure 6. The quantum circuit of the shift operator shifts the entire Read.

Only two auxiliary qubits are needed to shift the entire Read, regardless of the number of its residues. This fact shows that the quantum algorithm is scalable. As in the case of the compare operator, only CNOT quantum gates are used to construct the shift operator. Using the same quantum circuit, the reference DNA can also be shifted with reference to the Read.

*3.3 Outline of the hybrid quantum/classical system.*

The quantum algorithm described above comprises the compare and shift quantum operators that are applied consequently. Figure 7 shows the outline of the hybrid quantum/classical system. As shown in this figure, our quantum algorithm can be easily integrated into existing systems for DNA sequence alignment that use classical algorithms and computers.

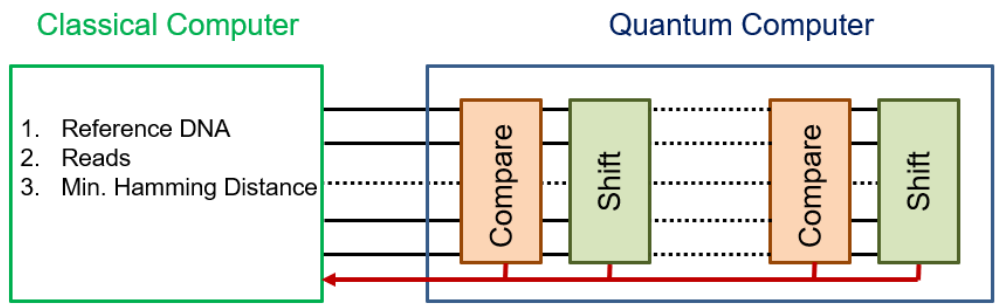

*Figure 7. Outline of the integration of our quantum algorithm into a classical computing system.*

In the outline of Figure 7, the classical computer holds the reference DNA and all the Reads. The reference DNA and the Reads are loaded to the quantum computer. The quantum computer compares and shifts the Read along the reference DNA and exports the Hamming distances to the classical computer. Finally, the quantum computer computes the minimum Hamming distance. Furthermore, as will be described in section 5, the Read states can be set in superposition, allowing the quantum computer to act as a quantum accelerator for the system.

**4. Execution of the quantum algorithm with Qiskit**

We used Qiskit to execute and validate our quantum algorithm on the quantum processing units and simulators provided by IBM Quantum, to show its correct operation. The steps of the execution of the quantum algorithm are as follows:

(1) Encode reference DNA and load it to the corresponding quantum register.

(2) Encode the Read to be aligned and load it to the corresponding quantum register.

(3) Compare the Read with the reference DNA.

(4) Shift by one site.

(5) Repeat from step (3) until all Read and reference DNA sequence comparisons have been completed.

Figure 8 shows the quantum circuits of the compare and shift operators executed using Qiskit. Figure 8(a) shows the quantum circuit for the compare operator and Figure 8(b) the shift operator encoded in Qiskit. The implementation of Figure 8 encodes a reference DNA with a length of six residues and a Read with a length of two residues.

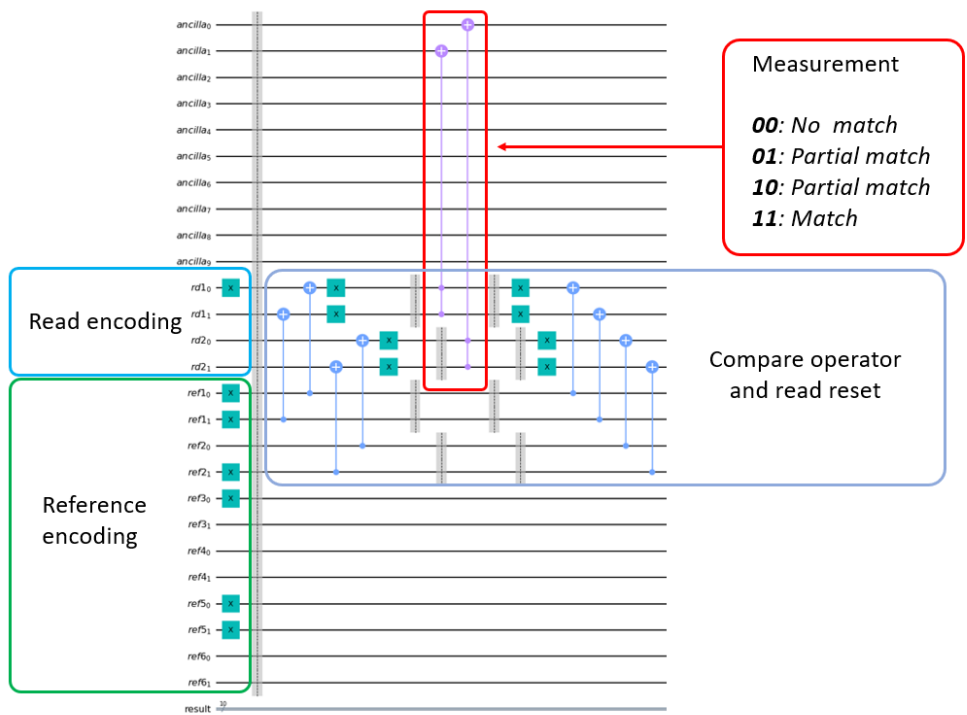

(a)

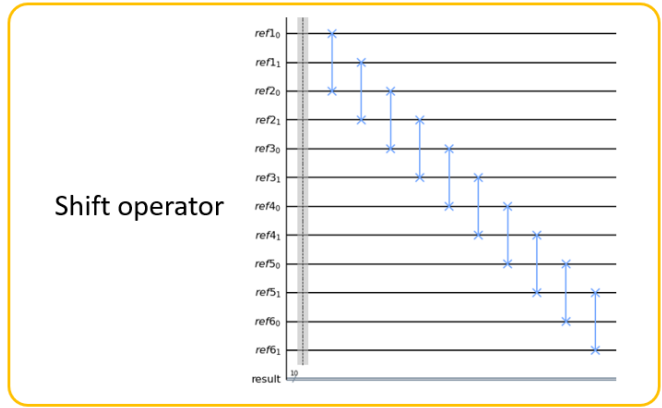

(b)

*Figure 8. Qiskit implementation for the quantum algorithm. (a) Encoding and compare operator (b) Encoding of the shift operator.*

Figure 9 shows the results of the quantum algorithm. The Read is aligned to the correct reference DNA site. The total number of qubits used is 26. The reference DNA sequence was encoded using 12 qubits and the Read sequence was encoded using 4 qubits. Ten ancilla qubits were used for the compare and shift operators.

```
Partial allignment after comparison 5
Read:CA
Ancilla qubits state: |01)

Alligned after comparison 3
Read:CA
Ancilla qubits state: |11)

Number of total qubits: 26
Number of ancilla qubits: 10
Number of qubits for the Read sequence: 4
Number of qubits for the Reference sequence: 12
```

| Reference | TGCATA | |
|---|---|---|
| Reference target | Read | Result |
| TG | CA | 00 |
| GC | CA | 00 |
| CA | CA | 11 |
| AT | CA | 00 |
| TA | CA | 01 |

*Figure 9. Results of the execution of the quantum algorithm on Qiskit. The Read "CA" was aligned to the correct site of the reference DNA "TGCATA". The Read is aligned at the third comparison (Result 11) as expected.*

Scaling up of the quantum algorithm is straightforward. We used the quantum circuits in Qiskit to implement the algorithm for a reference DNA with a length of ten residues and a Read with a length of four residues. Figure 10 shows the results of the quantum algorithm. Again, the Read is aligned to the correct reference DNA site. The total number of qubits used is 56. The reference DNA sequence was encoded using 20 qubits and the Read sequence was encoded using 8 qubits. The number of ancilla qubits used for the compare and shift operators is 28.

```
Alligned after comparison 5
Read:TAGG
Ancilla qubits state: |1111⟩

Partial allignment after comparison 1
Read:TAGG
Ancilla qubits state: |1110⟩

Number of total qubits: 56
Number of ancilla qubits: 28
Number of qubits for the Read sequence: 8
Number of qubits for the Reference sequence: 20
```

| Reference | TAGCTAGGCT | |
|---|---|---|
| Reference target | Read | Result |
| TAGC | TAGG | 1110 |
| AGCT | TAGG | 0000 |
| GCTA | TAGG | 0000 |
| CTAG | TAGG | 0001 |
| TAGG | TAGG | 1111 |
| AGGC | TAGG | 0010 |
| GGCT | TAGG | 0000 |

*Figure 10. Results of the execution of the quantum algorithm on Qiskit. The Read "TAGG" was aligned to the correct site of the reference DNA "TAGCTAGGCT". The Read is aligned at the fifth comparison (Result 1111) as expected.*

The quantum algorithm can align increased lengths of the reference DNA and the Read DNA as the number of available qubits increases. The number of qubits required to encode a residue is two. Thus, the total number of qubits used for the Read and reference encoding is given by:

$$qubits_{encoding} = 2 * (length_{read} + length_{reference}) \qquad (9)$$

The number of ancilla qubits, used for the comparison measurement is given by:

$$qubits_{ancilla} = length_{read} * (length_{reference} - length_{read} + 1) \qquad (10)$$

Hence, the total number of qubits required for the algorithm is:

$$qubits_{total} = qubits_{encoding} + qubits_{ancilla} \qquad (11)$$

## 5. Superposition of Reference and Read sequence quantum states

The quantum algorithm described above is scalable and can be easily incorporated into existing classical sequencing systems. The number of computations to align *n* Reads to the reference DNA is *n* and, therefore, the complexity of the algorithm is *O(n)*. To improve on the complexity and show the quantum advantage we developed an approach to set the Read and reference DNA states in superposition. Our approach is based on a modification of the Simon's quantum algorithm (Simon, 1997).

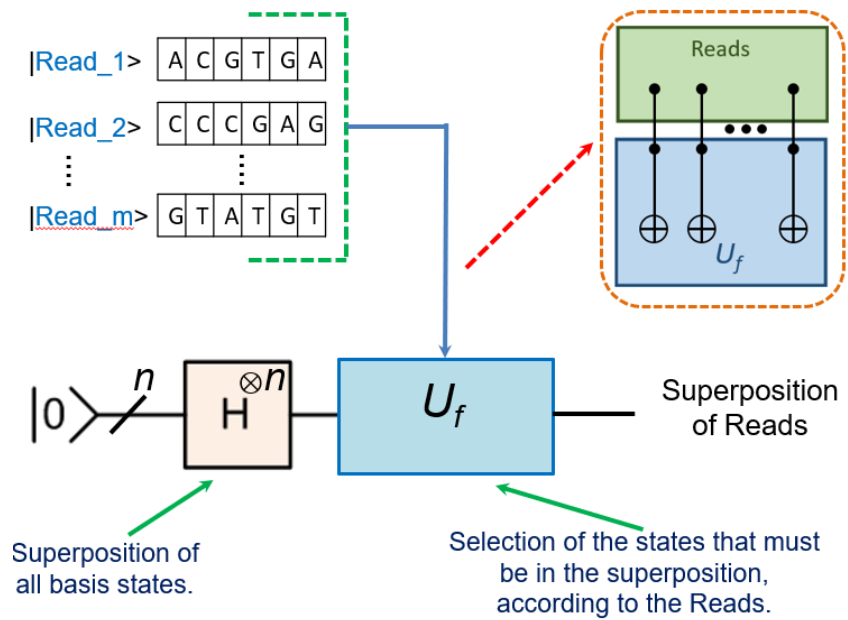

*Figure 11. Schematic description of the Read superposition problem.*

Figure 11 shows a schematic description of the process to set multiple Read states in superposition. The quantum computation starts with a quantum register with all its qubits in state $|0\rangle$. A superposition of all basis states is created by the action of the Hadamard gates. In our approach we use the quantum circuit $U_f$ to select the superposition states that correspond the specific $m$ Reads to be compared. The selection of states is controlled by the Read states using CCNOT quantum gates. The first control qubit is controlled by the Reads and the second control and target qubits belong to the quantum circuit $U_f$. At the end of the computation, the superposition of the Read states is obtained. The superposition of the reads will be compared and shifted using the quantum operators described in the previous sections.

Using the superposition of $m$ Reads the number of computations to align $n$ Reads is reduced to $n/m$ and the algorithmic complexity becomes $O\left(\frac{n}{m}\right)$.

Our approach to set the Read states in superposition is based on a modification of the Simon's quantum algorithm (Simon, 1997). The quantum circuit of Simon's quantum algorithm is shown in Figure 12.

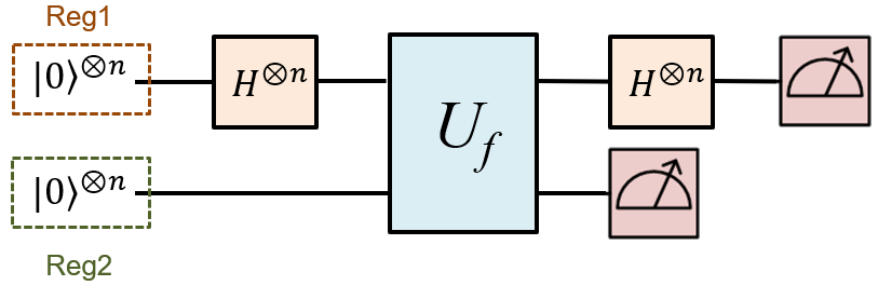

Figure 12. The quantum circuit of Simon's quantum algorithm.

In the original Simon's problem, a function $f : \{0, 1\}^n \to \{0, 1\}^n$ is given. There is also a hidden bit string $s = s_1, s_2, s_3, \cdots, s_n$ encoded using CNOT gates in the black box circuit $U_f$. The function is such that: $f(x) = f(y) \Leftrightarrow y = x \oplus s$. The aim of the quantum algorithm is to find the hidden string s. To set the specific Read states in superposition we modified the Simon's quantum algorithm, as shown in Figure 13.

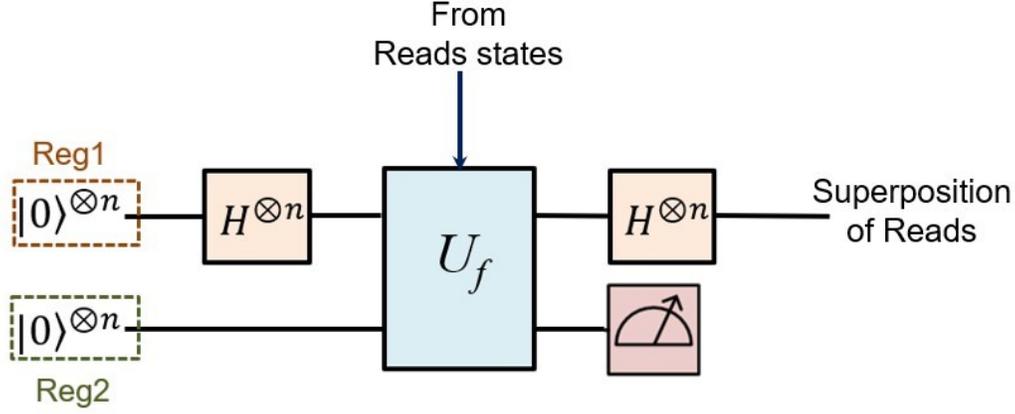

*Figure 13. The modified quantum circuit of Simon's quantum algorithm that sets the specific Read states in superposition.*

In the circuit of Figure 13, the specific Reads, the states of which are to be superposed, define the binary number *s*. To obtain this, CCNOT gates are used. The first control qubit is controlled by the Reads. Thus, the Read states activate and deactivate the CCNOT gates and define the binary number *s*. In our approach, the binary function *f(x)* maps the states that must be deleted from the superposition to the all-zero state and the states that must remain in non-zero states. The Hadamard gates that act on the register *Reg1* combined with the second control and target qubits in the circuit $U_f$ set the registers *Reg1* and *Reg2* in an entangled state. The combined state of the two registers right after the action of the second group of the Hadamard gates is:

$$|q\rangle = H^{\otimes n}\left(\frac{1}{\sqrt{2^n}}\sum_{x\in\{0,1\}^n}|x\rangle\right)\otimes |f(x)\rangle = \frac{1}{\sqrt{2^{n+1}}}\sum_{z\in\{0,1\}^n}\sum_{x\in\{0,1\}^n}(-1)^{x\cdot z}|z\rangle\otimes|f(x)\rangle \quad (12)$$

At this computation step, the register Reg2 contains the state that corresponds to the function *f(x)*, as defined by the Read states. Reg2 is measured. The two registers are entangled and the measurement of Reg2 defines the state of Reg1. If the measurement results in any non-zero state, then Reg1 contains the superposition of the Read states. If the measurement results in the all-zero state, then Reg1 contains the states that do not belong to the superposition and the quantum computation is repeated until a non-zero state is measured.

The reference DNA can also be divided into segments, and the segments can be set in superposition using the quantum circuit of Figure 13, with the first control qubit controlled by the reference DNA segments. In this manner, multiple Reads can be compared to multiple sites of the reference simultaneously.

Using a superposition of *m* Reads and a superposition of *k* reference DNA segments, the number of computations to align *n* Reads is reduced to $n/(m \cdot k)$ and the algorithmic complexity becomes $O\left(\frac{n}{m \cdot k}\right)$.

## 6. Conclusions and prospective medical applications

We presented a novel quantum algorithm for reference-guided DNA sequence alignment. The algorithm can be easily incorporated into existing classical sequence alignment systems because it is based on the quantization of the classical compare and shift operations. We simulated the quantum algorithm using Qiskit and showed its correct operation and its scalability. We described an approach based on Simon's quantum algorithm to set the Read and reference DNA states in superposition. We obtained an algorithmic complexity of $O\left(\frac{n}{m \cdot k}\right)$, that shows the quantum advantage of our algorithm. The quantum algorithm can align DNA sequences very fast. It can be applied in any assay performed with next generation sequencing (e.g. RNA-seq, ChIP-seq, Whole genome sequencing, Whole exome sequencing, Hi-C, etc.) which requires alignment of the DNA. Essentially, this algorithm can minimize the required time to analyze such assays, for example leading to the efficient identification of new biomarkers for disease treatment as a part of a CRISPR-Cas9 screening approach, which is used (and will often be used in the future) in personalized medicine.

## Acknowledgments

The views reflected in this article are the views of the author(s) and do not necessarily reflect the views of the global EY organization or its member firms.

## References

National Institute of General Medical Sciences [WWW Document], n.d. . National Institute of General Medical Sciences (NIGMS). URL https://nigms.nih.gov/

Berger, B., Waterman, M.S., Yu, Y. W., 2021. Levenshtein Distance, Sequence Comparison and Biological Database Search. IEEE Transactions on Information Theory, 67(6), 3287-3294, https://doi.org/10.1109/TIT.2020.2996543.

Camarillo-Guerrero, L. F., Almeida, A., Rangel-Pineros, G., Finn, R.D., Lawley, T.D., 2021. Massive expansion of human gut bacteriophage diversity. Cell 184(4), 1098-1109.e9. https://doi.org/10.1016/j.cell.2021.01.029.

Chang, W.L, Ren, T.T., Luo J., Feng, M., Guo, M., Lin, K.W., 2008. Quantum Algorithms for Biomolecular Solutions of the Satisfiability Problem on a Quantum Machine. IEEE Transactions on NanoBioscience, 7(3) 215-222. https://doi.org/10.1109/TNB.2008.2002286

Chao, J.Tang, F., Lei Xu, L., 2022. Developments in Algorithms for Sequence Alignment: A Review. Biomolecules, 12(4), 546; https://doi.org/10.3390/biom12040546

De, R., 2022. Quantum Accelerated Pattern Matching for Genome Sequencing in Complex RNA Secondary Structures. IEEE International Conference on Quantum Computing and Engineering (QCE), Broomfield, CO, 842-843. https://doi.org/10.1109/QCE53715.2022.00139.

Edgar, R.C., Taylor, J., Lin, V. et al. 2022. Petabase-scale sequence alignment catalyses viral discovery. Nature 602, 142–147. https://doi.org/10.1038/s41586-021-04332-2

Emani, P.S., Warrell, J., Anticevic, A. et al. 2021. Quantum computing at the frontiers of biological sciences. Nature Methods 18, 701–709. https://doi.org/10.1038/s41592-020-01004-3

Grover, L.K., 1996. A fast quantum mechanical algorithm for database search. STOC '96: Proceedings of the twenty-eighth annual ACM symposium on Theory of Computing, 212–219. https://doi.org/10.1145/237814.237866

Hollenberg, L.C.L., 2000. Fast quantum search algorithms in protein sequence comparisons: Quantum bioinformatics. Physical Review E, 62, 7532. https://doi.org/10.1103/PhysRevE.62.7532

Karafyllidis, I.G., 2005. Quantum computer simulator based on the circuit model of quantum computation. IEEE Transactions on Circuits and Systems I: Regular Papers, 52(8), 1590-1596. https://doi.org/10.1109/TCSI.2005.851999.

Karafyllidis, I.G., 2008. Quantum mechanical model for information transfer from DNA to Protein. Biosystems, 93(3), 191-198. https://doi.org/10.1016/j.biosystems.2008.04.002

Karafyllidis, I.G., 2012. Quantum gate circuit model of signal integration in bacterial quorum sensing. IEEE/ACM Transactions on Computational Biology and Bioinformatics, 9(2), 571-579. https://doi.org/10.1109/TCBB.2011.104

Krittanawong, C., Sun, T., Eyal, H., 2017. Big Data and Genome Editing Technology: A New Paradigm of Cardiovascular Genomics. Current Cardiology Reviews, 13(4), 301-304. https://doi.org/10.2174/1573403X13666170804152432

Li, H., Homer, N., 2010. A survey of sequence alignment algorithms for next-generation sequencing. Briefings in Bioinformatics, 11(5), 473–483, https://doi.org/10.1093/bib/bbq015

Mohseni, M., Omar, Y., Engel, G.S., Plenio M.B.(Eds), 2011. Quantum Effects in Biology. Cambridge University Press. ISBN-13 : 978-1107010802

Needleman, S.B, Wunsch, C.D., 1970. A general method applicable to the search for similarities in the amino acid sequence of two proteins. Journal of Molecular Biology, 48(3), 443-453. https://doi.org/10.1016/0022-2836(70)90057-4

Nielsen M.A, Chuang, I.L. Quantum Computation and Quantum Information. Cambridge University Press, 2011. ISBN-13 : 978-1107002173

Patel, A. 2001. Quantum database search can do without sorting. Physical Review A 64, 034303. https://doi.org/10.1103/PhysRevA.64.034303

Ponting, C. P., 2017. Big knowledge from big data in functional genomics. Emerging Topics in Life Sciences, 1 (3), 245–248. https://doi.org/10.1042/ETLS20170129

Rathee, M., Dilip, K., Rathee, R., 2021. DNA Fragment Assembly Using Quantum-Inspired Genetic Algorithm. Research Anthology on Advancements in Quantum Technology, 228-245. https://doi.org/10.4018/978-1-7998-8593-1.ch009

Shaikh, T.A., Ali, R., 2016. Quantum Computing in Big Data Analytics: A Survey. 2016 IEEE International Conference on Computer and Information Technology (CIT), Nadi, Fiji, 112-115. https://doi.org/10.1109/CIT.2016.79.

Simon, D.R., 1997. On the Power of Quantum Computation. SIAM Journal on Computing, 26 (5), 1474–1483. https://doi.org/10.1137/S0097539796298637

Smith, T.F., Waterman, M.S., 1981. Identification of common molecular subsequences. Journal of Molecular Biology, 147 (1) 195-197. https://doi.org/10.1016/0022-2836(81)90087-5

Stephen, Z.D., Lee, S.Y., Faghri, F., Campbell, R.H., Zhai, C., Efron, M.J., Iyer, R., Schatz, M.C., Sinha, S., Robinson G.E., 2015. Big Data: Astronomical or Genomical? PLoS Biology 13(7): e1002195. https://doi.org/10.1371/journal.pbio.1002195

Wilton, R., Szalay, A.S., 2022. Performance optimization in DNA short-read alignment, Bioinformatics, 38(8) 2081–2087. https://doi.org/10.1093/bioinformatics/btac066